# Electron-Irradiation Induced Nanocrystallization of Pb(II) in Silica Gels Prepared in High Magnetic Field

Takamasa Kaito[1*], Atsushi Mori[2†] and Chihiro Kaito[3]

*1. Graduate School of Engineering, Tokushima University, Tokushima 770-8506, Japan*

*2. Institute of Technology and Science, Tokushima University, Tokushima 770-8506, Japan*

*3. Department of Nanophysics in Frontier Project, Ritsumeikan University, 1-1-1 Nojihigashi, Kusatsu, Shiga 535-8577, Japan*

**Abstract:** In a previous study, structure of silica gels prepared in a high magnetic field was investigated. While a direct application of such anisotropic silica gels is for an optical anisotropic medium possessing chemical resistance, we show here their possibility of medium in materials processing. In this direction, for example, silica hydrogels have so far been used as media of crystal growth. In this paper, as opposed to the soft-wet state, dried silica gels have been investigated. We have found that lead (II) nanocrystallites were formed induced by electron irradiation to lead (II)-doped dried silica gels prepared in a high magnetic field such as $B$ = 10 T. Hydrogels made from a sodium metasilicate solution doped with lead (II) acetate were prepared. The dried specimens were irradiated by electrons in a transmission electron microscope environment. Electron diffraction patterns indicated the crystallinity of lead (II) nanocrystallites depending on $B$. An advantage of this processing technique is that the crystallinity can be controlled through the strength of magnetic field $B$ applied during gel preparation. Specific skills are not required to control the strength of magnetic field.

## 1. Introduction

Electron irradiation (EI) is a potential treatment of materials. EI treatment of polymers began early in 1950. Structure of collagen was changed by EI [1]. EI treatments were not limited to polymers. In 1968, a copper foil doped with cobalt was irradiated by electrons and it was observed that precipitates lost their coherency [2, 3]. Nanofabrications were performed using EI. Crystallization or nanocrystallization of iron based amorphous alloy induced by EI was reported [4-9]. Metallic glass of Neodium alloy (FeNdB) was nanocrystallized induced by EI [10-12]. Nanocrystallization of CuZrTi metallic glass by EI was investigated [13].

*Present address: KRI Inc.

†**Corresponding author:** Atsushi Mori, Associate Professor / Dr. (Eng.), research field: Soft matter. E-mail: atsushimori@tokushima-u.ac.jp

The subject of this paper is EI-induced lead (II) nanocrystallization in dried silica gels. The matrix is not an amorphous metallic alloy but an insulator. Crystalline silicon nanodots were formed in an amorphous silica film by EI [14]. Si-O-C freestanding nanowires were grown by EI form a tetraethylorthosilicate (TEOS) precursor [15]. Also, using TEOS precursors nanostructures were grown by electron beam-induced deposition of Pt [16]. By EI zinc nonocrystal islands were formed in an $SiO_2$ layer [17]. Moreover, periodic zinc nanocrystal arrays were formed in an amorphous ZnSiO layer by EI [18]. In those studies the matrix is a non-crystalline insulator (oxide), which is same as in the present study.

In 2006, we grew lead (II) bromide ($PbBr_2$) crystals using silica hydrogels as media of a crystal growth [19, 20]. The silica gels were made from sodium metasilicate ($Na_2SiO_3$) aqueous solution by adding a concentrated acetic acid solution. As a source of $Pb^{++}$ ions a lead (II) nitrite $(Pb(NO_3)_2)$ solution was added. This mixed solution was settled in a magnetic field such as $B = 5$ T to prepare the silica hydrogels. We observed an ordered array of $PbBr_2$ nanocrystallites with their crystallographic axes aligned along the direction of the magnetic field. It is suggested from the fact that no magnetic field was applied during the crystal growth that a structural anisotropy was formed in the silica hydrogels. To detect the anisotropy in the silica gels, in Ref. [21] we measured birefringence and performed scanning microscopic light scattering (SMILS) (developed by Furukawa and his coworkers [22]). At that time, we detected negative birefringence of magnitude of $10^{-6}$ with the direction of the magnetic fields being as the optic axis. We compared results of birefringence measurement of samples prepared in borosilicate glass cell and quartz one [23]. It was found that whereas the negative birefringence was obtained for the borosilicate glass cells, a positive birefringence was observed for the quartz cells. Birefringence of silica hydrogels has recently been reinvestigated using quartz cells [24]; the conclusion of the negative birefringence has been overturned. In addition, improvement of the equipment of the Sènarmont measurement has now been in progress [25]. Statistical accuracy will be increased thereby. Nevertheless, the results of SMILS did definitely indicate ordering. The width of distribution of relaxation time of autocorrelation function of the electric fields of scattered light was smaller for samples exhibiting birefringence than for ones exhibiting no birefringence. Even if one suspects the results of birefringence, we can say that the samples are classified into two, one has a narrow size distribution of the gel network and the other a wide one. It follows that an application of magnetic field during silica gel preparation brings certain order in the gel network.

In this paper, we report formation of Pb(II) nanocrystallites in dried silica gels induced by EI. The silica hydrogels were prepared in a high magnetic field such as $B = 10$ T with $Na_2SiO_3$ aqueous solution being as a starting material such as done previously [19, 20].

## 2. Materials and Methods

At first, 50 g sodium metasilicate $(Na_2SiO_3 \cdot 9H_2O)$ was resolved in 103 ml distilled water as described in a previous paper [19]. Then, 8 ml of this solution was diluted with 8 ml distilled water. For gellation of the $Na_2SiO_3$ aqueous solution, we made this solution strongly aciditic by adding a concentrated acetic acid solution. 16 ml concentrated acetic acid aqueous solution was added under stirring. Then, 6.4 ml of 1M $Pb(CH_3COO)_2$ aqueous solution was added. An advantage of the use of $Pb(CH_3COO)_2$ over $Pb(NO_3)_2$ is that one does not need to pay special attention to avoid formation of cloudiness at the addition of source of Pb(II). In the case of $Pb(NO_3)_2$, we should pay attention not to make inhomogeneity in mixing of nitrite and acetate. This mixed solution was stirred for 2 hours. And then, the mixture was separated into test tubes and sealed. After that, they were put in equipment of a high magnetic field, a superconducting magnet (JMTD-10100M, Japan Superconductor Technology, Inc.). Three different magnetic fields (10, 8, and 4 T) were applied by setting the samples at different positions. The samples were gellated by settlement for a week.

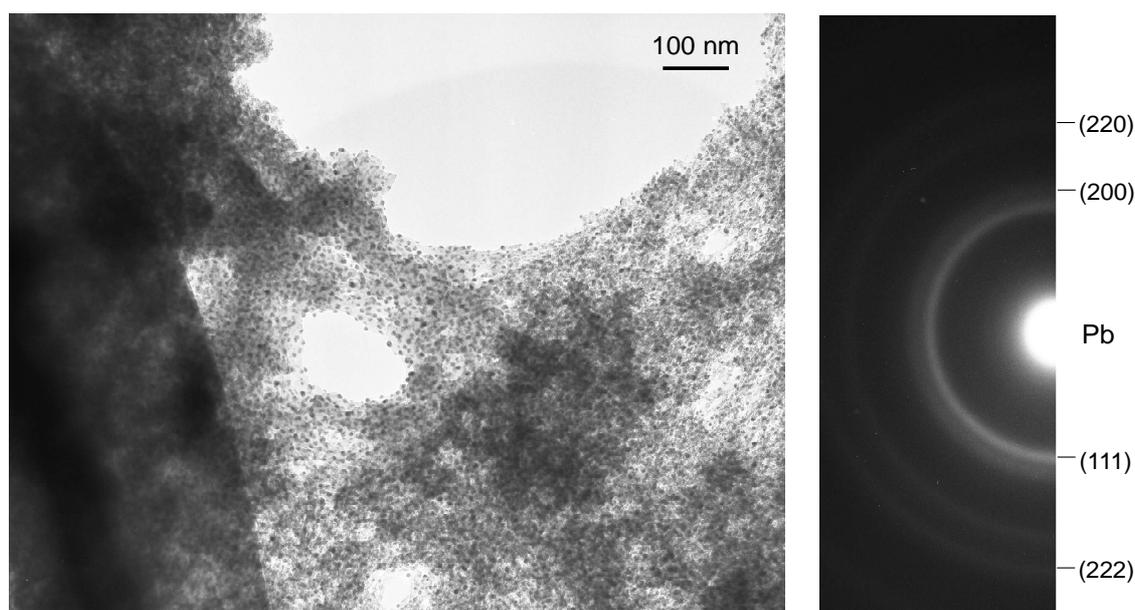

**Fig. 1 A result of TEM observation for a sample prepared under 10 T: (a) TEM image, (b) ED pattern. A large aggregate of precipitates is seen in the right-below from the center in (a). Debye rings are indexed in (b).**

Difference between uses of $Pb(CH_3COO)_2$ and $Pb(NH_3)_2$ as the source of Pb(II) was not detected in birefringence measurements. The high transparency of silica hydrogels brought low adsorption and scattering of light was neglected in both cases. So, the materials were suggested to be essentially the same regarding the structural anisotropy. We did not make efforts to detect the minor difference; in the present stage, we wish to regard that only an easiness in material preparation was bright.

Dried silica gels, which we obtained by drying the hydrogels for a year in sealed test tubes, were irradiated by electrons at 100 kV for minutes in a transmission electron microscope (TEM) (Hitachi H-9000NAR) environment. This condition was the same as those for previous TEM observations [19]. One 10 T sample and two 4 T samples were examined.

## 3. Results and Discussion

Figure 1 is a result of TEM observation of the sample prepared under 10 T. A TEM image and an electron diffraction (ED) pattern are shown. Rings indicating crystallinity of face-centered cubic (fcc) Pb(II) are seen in the ED pattern. We have identified that these rings are of (111), (200), (220), and (222) diffraction. We confirm nanocrystallites of a few tens of nanometers in aggregates in the TEM image. Formation of Pb(II) nanocrystallites is suggested. We speculate that this nanocrystallization was induced by EI. There remains, however, a possibility that nanocrystallites were formed during drying. To exclude this possibility one should perform another experiment without EI. However, EI induced nanocrystallization is likely since there were a lot of reports on EI-induced nanocrystallization as mentioned in Section 1.

Before proceeding to the next sample, distinction between previous work [19, 20] and the present one is described. The previous study was organized to grow $PbBr_2$ crystals through a reaction between $Pb^{++}$ in $Pb(NO_3)_2$ and $Br^-$ in KBr. While the former was doped to the silica hydrogels, the latter was provided in an aqueous solution poured on the gels. It is a key that the material transportation in the gels is followed by the reaction. On the other hand, such mechanisms are absent. $Pb^{++}$ ions can aggregate prior to EI. It is a quite reasonable scenario that $Pb^{++}$ is reduced to Pb by EI and then precipitates.

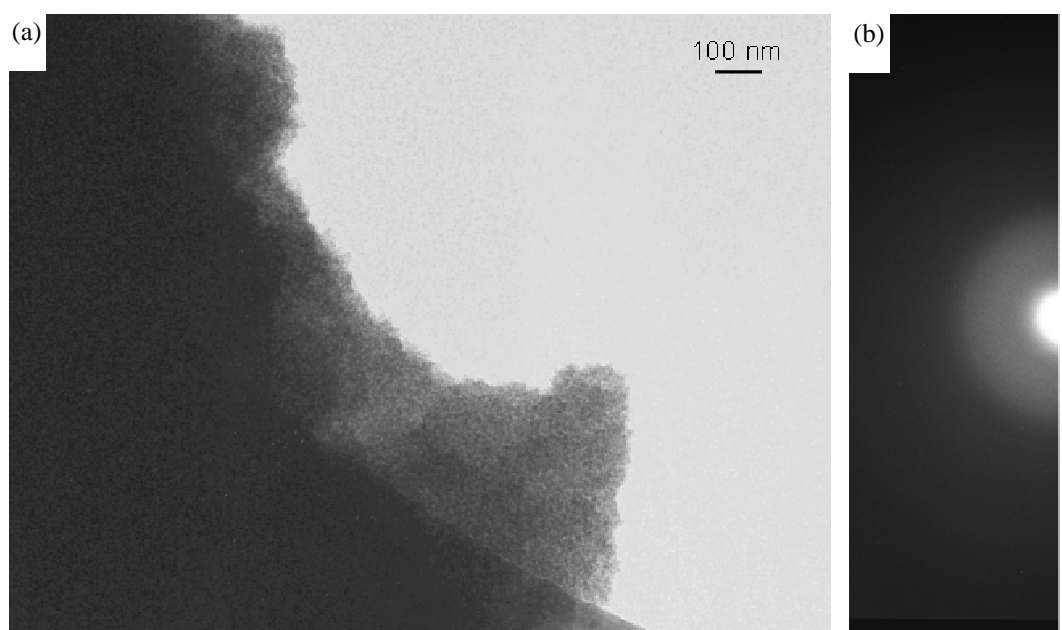

**Fig. 2 A result of TEM observation for one of samples prepared under 4 T: (a) TEM image, (b) ED pattern. Besides small ones, unlike in Fig. 1(a) aggregates are not grown largely in (a). ED pattern is breadened in (b).**

Let us look into a result of TEM observation of one of the samples prepared under 4 T (Fig. 2). In the ED pattern the Debye ring is broad and no diffraction patterns from crystalline lattices are suggested. Therefore, the TEM image is suggested to be that of an amorphous aggregate precipitated or nucleated on an inside surface of a macropore (the right white region in the TEM image is a macropore and one sees an aggregate at the left border of this region). Structure inside the precipitate is not very clearly shown, unfortunately, because of a low sharpness of the image. The sharpness may be improved by image processing. To detect an onset of crystallization by doing so may be possible, but this is not a present concern.

Let us proceed to the other sample prepared under 4 T (Fig. 3). Although the Debye rings are a little broadened, as for the 10 T sample, rings indicating the crystallinity are seen. We have also identified (111) and (200) rings. A more broadened rings are seen, but we have not identified. It is sufficient for present purpose to point out that the crystallinity of this sample is certainly worse than that for the 10 T sample. Aggregates of nanocrystallites of a few tens of nanometers are observed on inside surfaces of a macropore (the top white region in the TEM image is a macropore and one finds aggregated nanocrystallites at the left and bottm-right borders -- the aggregated nanocrystallites seemingly grew into the macropore at the center).

Comparing Figs. 1(a) and 3(a) we find that the rings are sharp for the 10T sample as has been mentioned. We conjecture that higher order in the gel network formed in a higher magnetic field brought this result. Formation of an amorphous precipitate in the 4T sample gives a support to a trend that the crystalline order increases with the strength of applied magnetic field. One can state that order induced by the magnetic field in the gels affects certainly the crystallinity of the nanocrystallites without inferring when the ordering occurs. The fact that whether the materials are crystalline of amorphous can be controlled by the strength of applied magnetic field is advantageous because no specific skills are necessary in controlling the strength of magnetic field. In addition, we note that all treatments were done in room temperature. That is, no thermal treatments are required as in a recent study of EI-induced nanocrystallization of Ni [26].

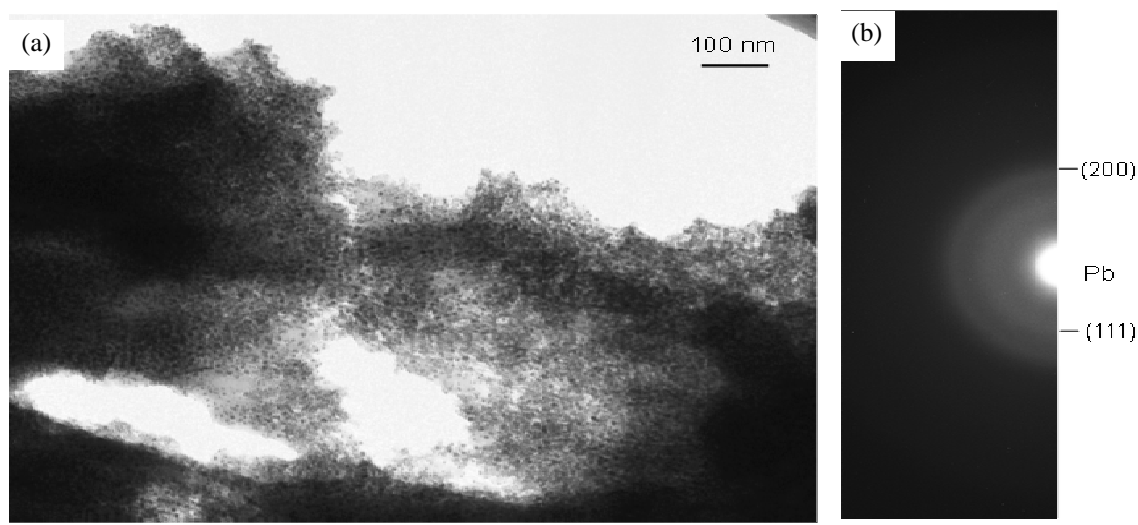

**Fig.3 A result of TEM observation for the other sample prepared under 4 T: (a) TEM image, (b) ED pattern. Aggregated precipitates as elongated from the right-bottom wall to the center are seen in (a). Debye rings can be indexed in (b).**

The size of nanocrystallites of a few tens of nanometers is comparable to that of the short axis of nanocrystallites observed in the crystal growth of $PbBr_2$ in silica hydrogels prepared in a magnetic field [19]. This size may be on the same order of one of the characteristic lengths of the magnetically ordered network of silica hydrogels. This size is also comparable to the pore size of the small one of bimordal pore size distribution of silica xerogels as reported in Refs. [27, 28].

The macropores of a few micrometers seen in Figs. 1(b), 2(b), and 3(b) correspond to the large one in the bimordal pore size distribution. We speculate that such macropores form in the drying process and the pore walls are made of bundled silica polymers. In this respect, the dried silica gels under investigation are xerogels; i.e., the structure of the dried gels was changed from that of the wet gels. Nevertheless, ordered network structure induced by a magnetic field must remain although the pore size must change a little. During EI-induced nanocrystallization, those orders affect and bring results such as obtained here.

For application of this method, the crystallinity of nanocrystallites should be quantified as a function of the strength of applied magnetic field. The crystallinity can be quantified such as based on the width of the Debye rings in ED patterns as a function of the strength of magnetic field. Also, condition of preparation of silica gels, other than the magnetic field, should be optimized. In the preliminary stage, we irradiated the sample by electrons in a TEM environment for TEM observations as done previously [19]. The condition of EI as well as drying of gels should also be optimized. We wish to postpone such studies as future researches.

In some future, Pb(II) can be replaced with semiconductor or metal. Pb(II) as well as semiconductor/metal nanocrystallite dispersed silica gels can be used in applications such as quantum size effect devices and in nanophotonics. In some cases nanocrystallites alone may be utilzed, in some cases silica gels including nanocrystallites may be useful. In particular, we expect improved performance of plasmonic penomena for noble metal dispersed silica gels. Selection of the source of the semiconductor/metal may be important factor for safe in use. Identification of ordered networks in wet silica hydrogels is underway.

## 4. Summary

We have examined electron irradiation to Pb(II)-doped silica gels prepared in a magnetic field. It was remarkable that the crystallinity of nanocrystallites formed by the electron irradiation could be controlled through the strength of the magnetic field. Potential to open a new field of techniques has been pointed out. Some developments in some future based on this phenomena have also been discussed.

## Acknowledgment

It is acknowledged that Prof. T. Ichitsubo kindly provided an environment of high magnetic field.

## References

[1] Perron, R. R. and Wright, B. A. Alteration of Collagen Structure by Irradiation with Electrons. *Nature* **1950,** *166*, 863-854.

[2] Brown, L. M., Woolhouse, G. R. and Valdre, U., Radiation-induced Coherency Loss in a Cu-Co Alloy. *Philos. Mag.* **1968,** *17*, 781-789.

[3] Woolhouse, G. R. Loss of Precipitate Coherency by Electron Irradiation in the High Voltage Electron Microscope. *Nature* **1968,** *220*, 573-574.

[4] Nagase, T., Umakoshi, Y. and Sumida, N. Effect of Electron Irradiation on the Phase Stability of Fe-9Zr-3B alloy. *Mater. Sci. Eng. A* **2002,** *323*, 218-225.

[5] Nagase, T., Umakoshi, Y. and Sumida, N. Formation of Nanocrystalline Structure during Irradiation Induced Crystallization in Amorphous Fe-Zr-B Alloys.. *Sci. Technol. Adv. Mater.* **2002,** *3*, 119-128.

[6] Nagase, T and Umakoshi, Y. Effect of Irradiation Temperature on the Electron Irradiation induced Nanocrystallization Behavior in Fe$_{88.0}$Zr$_{9.0}$B$_{3.0}$ Amorphous Alloy. *Mater. Sci. Eng. A* **2003,** *347*, 136-144.

[7] Nagase, T. and Umakoshi, Y. Electron Irradiation Induced Nanocrystallization Behavior in Fe$_{71}$Zr$_9$B$_{20}$ Metallic Glass. *Mater. Trans.* **2005,** *46*, 608-615.

[8] Nagase, T., Nino, A. and Umakoshi, Y. Phase Stability of an Amorphous Phase Against Electron Irradiation Induced Crystallization in Fe-Based Metallic Glasses. *Mater. Trans.* **2007,** *48*, 1340-1349.

[9] Qin, W, Nagase, T and Umakoshi, Y. Electron Irradiation-induced Nanocrystallization of Amorphous Fe$_{85}$B$_{15}$ alloy: Evidence for Athermal Nature. *Acta Mater.* **2009,** *57*, 1300-1307.

[10] Nino, A., Nagase, T. and Umakoshi, Y. Electron Irradiation Induced Nano-Crystallization in Fe$_{77}$Nd$_{4.5}$B$_{18.5}$ Metallic Glass. *Mater. Trans.* **2005,** *46*, 1814-1819.

[11] Nino, A., Nagase, T and Umakoshi, Y. Electron Irradiation Induced Phase Transformation in Fe-Nd-B Alloys. *Mater. Trans.* **2007,** *48*, 1659-1664.

[12] Nagase, T., Nino, A and Umakoshi, Y. Nano-Crystallization and Stability of an Amorphous Phase in Fe-Nd-B Alloy under 2.0 MeV Electron Irradiation. *Mater. Trans.* **2008,** *49*, 265-274.

[13] Xie, G., Zhang, Q., Louzgguine-Luzgin, D. V., Zhang, W. and Inoue, A. Nanocrystallization of Cu$_{50}$Zr$_{45}$Ti$_5$ Metallic Glass Induced by Electron Irradiation. *Mater. Trans.* **2006,** *47*, 1930-1933.

[14] Du, W.-W., Takeguchi, M., Tanaka, M. and Furuya, K. Formation of Crystalline Si Nanodots in SiO$_2$ Films by Electron Irradiation. *Appl. Phys. Lett.* **2003,** *82*, 1108-1110.

[15] Frabboni, S., Gazzadi, G. C. and Spessot, A. Transmission Electron Microscopy Characterization and Sculpting of Sub-1 nm Si-O-C Freestanding Nanowires Grown by Electron Beam Induced Deposition. *Appl. Phys. Lett.* **2006,** *89*, 113108.

[16] Gazzadi, G. C., Frabboni, S. and Menozzi, C. Suspended Nanostructures Grown by Electron Beam-induced Deposition of Pt and TEOS Precursors. *Nanotechnol.* **2007,** *18*, 445709.

[17] Kim, T. W., Shin, J. W., Lee, J. Y., Jung, J. H., Lee, J. W., Choi, W. K. and Jin, S. Electron-beam-induced Formation of Zn Nanocrystal Islands in a SiO$_2$ Layer. *Appl. Phys. Lett.* **2007,** *90*, 051915.

[18] Shin, J. W., Lee, J. Y., No, Y. S., Kim, T. W., Choi, W. K. and Jin, S. The Formation Mechanism of Periodic Zn Nanocrystal Array Embedded in an Amorphous Layer by Rapid Electron Beam Irradiation. *Nanotechnol.* **2008,** *19*, 295303.

[19] Kaito, T., Yanagiya, S.-i., Mori, A., Kurumada, M., Kaito, C. and Inoue, T. Effect of Magnetic Field on the Gel Growth of PbBr$_2$. *J. Cryst. Growth* **2006,** *289*, 275-277.

[20] Kaito, T., Yanagiya, S.-i., Mori, A., Kurumada, M., Kaito, C. and Inoue, T. Characteristic Nanocrystallite Growth of PbBr$_2$ in a Magnetic Field in Gel. *J. Cryst. Growth* **2006,** *294*, 407-410.

[21] Mori, A., Kaito, T. and Furukawa, H. Structural Anisotropy of Silica Hydrogels Prepared under Magnetic Field. *Mater. Lett.* **2008,** *62*, 3459-3461.

[22] Furukawa, H., Horie, K., Nozaki, R. and Okada, M. Swelling-induced Modulation of Static and Dynamic

Fluctuations in Polyacrylamide Gels Observe by Scanning Microscopic Light Scattering. *Phys. Rev. E* **2003,** *68,* 031406.

[23] Tomita, R., Mori, A., Yamato, M., Futrukawa, H. and Takahashi, K. Structural Anisotropy of Silica Hydrogels Prepared under Magnetic Field. *World J. Eng.* **2010,** *7 (Suppl. 2)*, 518-519.

[24] Mori, A., Kaito, T., Furukawa, H., Yamato, M. and Takahashi, K. Birefringence of Silica Hydrogels Prepared under High Magnetic Fields Reinvestigated. *Mater. Res. Exp.* **2014,** *1*, 045202.

[25] Mori, A. and Tomita, R. Semi-Automated Sènermont Method for Measurement of Small Retardation. *Instr. Sci. Technol.* **2015,** *43,* in press (DOI: 10.1080/10739149.2014.1003072).

[26] del Angel, P., Rodriguez-Hernandez, J. H., Garcia-Borquez, A. and de la Fuente, J. A. M. Nucleation and Growth of $Ni^0$ Nanoparticles and Thin Films by TEM Electron Irradiation. *Catalys. Today* **2013,** *212*, 194-200.

[27] Takahashi, R., Nakanishi, K. and Sogo, N. Effect of Aging and Solvent Exchange on Pore Structure of Silica Gels with Interconnected Macropores. *J. Non-Cryst. Solids* **1995,** *189*, 66-76.

[28] Hoffmann, H., Meyer, M. and Zeitler, U. Control of Morphology inside the Mesoporous Gelstructure in Silica-gels. *Colloids Surf. A* **2006,** *291*, 117-127.